\documentclass[twocolumn]{aastex631}

\def\spose#1{\hbox to 0pt{#1\hss}}
\def\simlt{\mathrel{\spose{\lower 3pt\hbox{$\mathchar"218$}}
     \raise 2.0pt\hbox{$\mathchar"13C$}}}
\def\simgt{\mathrel{\spose{\lower 3pt\hbox{$\mathchar"218$}}
     \raise 2.0pt\hbox{$\mathchar"13E$}}}

\begin{document}


\title{Conditions for Changing-Look AGNs from Accretion Disk-Induced Tidal Disruption Events}

\author[0000-0002-8614-8721]{Yihan Wang}
\affiliation{Nevada Center for Astrophysics, University of Nevada, Las Vegas, NV 89154, USA}
\affiliation{Department of Physics and Astronomy, University of Nevada Las Vegas, Las Vegas, NV 89154, USA}

\author[0000-0002-3168-0139]{Matthew J. Graham}
\affiliation{Division of Physics, Mathematics and Astronomy, California Institute of Technology, Pasadena, CA 91125, USA}

\author[0000-0002-5956-851X]{K.E. Saavik Ford}
\affiliation{Center for Computational Astrophysics, Flatiron Institute, 
162 5th Ave, New York, NY 10010, USA}
\affiliation{Department of Astrophysics, American Museum of Natural History, New York, NY 10024, USA}
\affiliation{Department of Science, BMCC, City University of New York, New York, NY 10007, USA}

\author[0000-0002-9726-0508]{Barry McKernan}
\affiliation{Center for Computational Astrophysics, Flatiron Institute, 
162 5th Ave, New York, NY 10010, USA}
\affiliation{Department of Astrophysics, American Museum of Natural History, New York, NY 10024, USA}
\affiliation{Department of Science, BMCC, City University of New York, New York, NY 10007, USA}

\author[0000-0003-2012-5217]{Taeho Ryu}
\affiliation{Max Planck Institute for Astrophysics, Karl-Schwarzschild-Str.~1, 85748 Garching, Germany}

\author[0000-0003-2686-9241]{Daniel Stern}
\affiliation{Jet Propulsion Laboratory, California Institute of Technology, 4800 Oak Grove Drive, Pasadena, CA 91109, USA}



\begin{abstract}
The phenomenon of changing-look (CL) behavior in active galactic nuclei (AGN) is characterized by dramatic changes in luminosity and/or emission line profiles over relatively short periods, ranging from months to years. The origin of CL-AGNs remains a mystery, but one proposed explanation involves the response of the inner AGN disk to tidal disruption events (TDEs) around the supermassive black hole (SMBH). In this Letter, we calculate the predicted frequency of AGN TDEs as a function of SMBH mass and compare the results to the observed CL-AGN distribution. We find that if the fraction of CL-AGNs caused by AGN-TDEs is high, then: (1) most SMBHs in CL-AGN are near maximal spin, with the dimensionless spin parameter $a>0.9$; (2) AGN inner disks have a high surface density ($\geq 10^{7}\, {\rm g\, cm^{-2}}$); (3) typical AGN lifetimes are $\sim 10$-$100$ Myr; and (4) a nuclear star cluster initial mass function (IMF) that scales as $\sim m_*^{-1.6}$ is preferred. Future observations of CL-AGN will help constrain the fraction of CL-AGNs caused by AGN-TDEs, SMBH spins, AGN lifetimes, and the nuclear star cluster IMF. 



\end{abstract}




\section{Introduction} \label{sec:intro}

Active Galactic Nuclei (AGN) are powered by accretion onto supermassive black holes (SMBHs) and are characterized by broad-band multi-wavelength emission as well as variability across a range of timescales \citep[e.g.,][]{Heckman14}. Recently, a new category of AGN variability has been identified: a phenomenon known as changing-look (CL) AGNs involving dramatic changes in continuum emission \citep[e.g.,][]{LaMassa15, Ruan2016} and/or emission line profile \citep[e.g.,][]{Runco16, Graham2020}. This phenomenon is inconsistent with inner disk viscous accretion timescales \citep[e.g.,][]{Stern2018} and is seen across a range of redshifts \citep{Ross20}. Proposed explanations for CL-AGN include obscuration along the AGN sightline \citep{Nenkova2008, Shapovalova2010, Elitzur2012}, sudden changes in accretion rate \citep{Eracleous1995, Elitzur2014, Yang2018, Wang2018, WangJM2024}, abrupt changes in the temperature structure of the inner accretion disk \citep{Ross2018, Stern2018}, or tidal disruption events \citep[TDEs;][]{Merloni2015, Blanchard2017}.

Although obscuration is clearly responsible for some CL-AGN phenomena, particularly events identified at X-ray energies \citep[e.g.,][]{Risaliti02, Rivers15}, evidence supports intrinsic changes for most events selected at other wavelengths. \citet{Mathur2018} report complex multi-band recovery inconsistent with obscuration in the CL-AGN Mrk~590, while strong month-long timescale infrared variability cannot be due to obscuration of the parsec-scale dusty torus \citep{Sheng2017, Stern2018}. The low polarization levels seen in CL-AGN also argues against scattering and obscuration \citep{Hutsemekers2019}. Finally, the lagged response between optical emission from the accretion disk and infrared emission from the extended dusty torus implies that CL-AGN activity is due to intrinsic changes to the accretion disk luminosity rather than obscuration \citep{Kynoch2019}. 

Intrinsic changes in bolometric luminosity significantly affect the X-ray spectra and the broad-line regions (BLR) in CL-AGN. These changes could be related to TDEs or the spontaneous behavior of the accretion flow near a black hole. While TDEs may explain the intrinsic changes of the central engine, they are statistically unlikely due to the low TDE rate ($10^{-6} - 10^{-4}\, {\rm yr}^{-1}$ per galaxy), which is inconsistent with CL-AGN activities ($10^{-4}-10^{-3}$ per AGN yr$^{-1}$) if the TDE rate in quiescent galaxies is similar to the TDE rate in AGNs.

Recently, however, several theoretical studies have inferred that TDEs should occur much more frequently in AGNs \citep{Wang2024, WangMY2024, Kaur2024}, where they occur due to disk-star interactions, compared to quiescent galaxies, where relaxation processes dominate. These AGN-TDEs can result from orbital coincidence \citep{McK20}, orbital capture from the nuclear star cluster \citep[NSC;][]{MacLeod2020, Fabj2020, Generozov2023, Wang2023b}, and star formation \citep{Yuri07}. Embedded stars on retrograde orbits are excellent candidates for AGN-TDEs, particularly in the early AGN lifetime \citep[e.g.,][]{McK22, Wang2024}. 

Another challenge for the TDE-induced CL-AGN scenario is that CL-AGNs are commonly found in AGNs with more massive SMBHs (i.e., $\simgt 10^8 M_\odot$), where the tidal disruption radius of $\sim 1 M_\odot$ main sequence stars is within the event horizon of the SMBH. TDEs should not occur in such AGNs, and no intrinsic central engine changes are expected. However, the physical tidal disruption radius depends on the star’s density profile and mass \citep{Ryu2020a, Ryu2020b} and the event horizon is smaller in the equatorial plane of spinning SMBHs. Thus, TDEs can still occur in high-mass spinning AGNs if the disrupted star is massive and/or has a high density (e.g., white dwarfs). Although more massive stars are less common due to the shape of the initial mass function (IMF) \citep[e.g., Salpeter IMF;][]{Salpeter1955, Kroupa2002}, massive stars may be relaxed or transported to the vicinity of the SMBH due to mass segregation and disk transportation \citep{Zhou2022,wu2024,Zhou2024}. Indeed, \citet{Hinkle2024} recently identified three extreme nuclear transients that they suggest are associated with the tidal disruption of intermediate mass ($\sim 3-10\, M_{\odot}$) stars.

Based on the intriguing nature of this phenomenon, in this Letter we investigate the parameters of NSC and accretion disk models required in order for the rate of TDE-induced CL-AGN to align with published CL-AGN observations. Section~\ref{sec:method} describes the models of AGN-TDEs, AGN disks, and NSCs. Section~\ref{sec:results} compares the inferred AGN-TDE rates with the CL-AGN rate. We further discuss TDE-induced CL-AGNs and summarize our results in Section~\ref{sec:con}.

\section{Methods} \label{sec:method}

\subsection{Disk-captured TDEs}\label{sec:disk-tde}

\begin{figure}
    \includegraphics[width=\columnwidth]{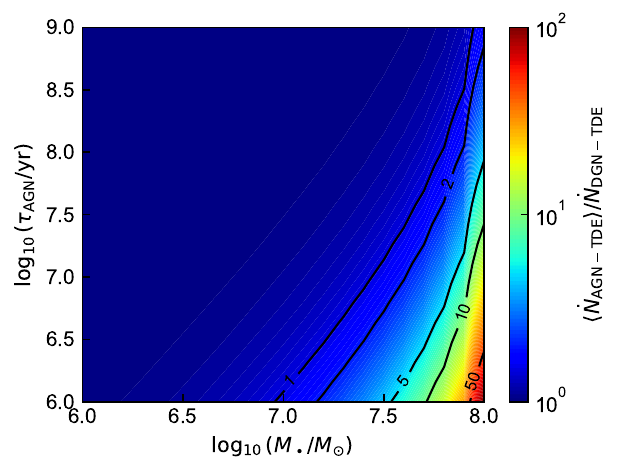}
    \caption{The ratio between the averaged AGN TDE rate ($\langle\dot{N}_{\rm AGN-TDE}\rangle$) caused by disk-star interactions during the AGN lifespan ($\tau_{\rm AGN}$) and the dormant galactic nuclei TDE rate ($\dot{N}_{\rm DGN-TDE}$) due to two-body relaxations. Massive AGNs with short-lived disks exhibit a significantly higher TDE rate compared to dormant galaxies.}
    \label{fig:enhanced}
\end{figure}

Inclined stars from the NSC interact with the AGN disk mainly due to aerodynamic drag. This interaction alters the stars' orbital properties, eventually resulting in their capture by the disk \citep[][]{artymowicz1993, Fabj2020, Generozov2023, Wang2023b}. This process preserves the quantity $L\cos^2(I/2)$ \citep{Wang2023b}, where $L$ is the specific angular momentum and $I$ is the orbital inclination. The resulting captured semi-major axis is given by
\begin{eqnarray}
{a_{\rm cap}} = a_0(1-e^2_0)\cos^4(I_0/2)\,,
\end{eqnarray}
when the final orbit is circularized and co-planar, where $a_0$, $e_0$, and $I_0$ are the initial semi-major axis, eccentricity, and inclination angle, respectively. Stars with initial inclinations exceeding a critical value,
\begin{eqnarray}
I_{\rm TDE} = 2\arccos\left(\left(\frac{r_{\rm t}}{a_0(1-e_0^2)}\right)^{1/4}\right)
\end{eqnarray}
will be tidally disrupted by the SMBH during the disk capture process, where $r_{\rm t}\sim (\frac{3M_\bullet}{m_*})^{1/3}R_*$ is the tidal disruption radius of a main sequence star with mass $m_*$ and radius $R_*$, and $M_\bullet$ is the SMBH mass. For most of the cases where $r_t\ll a_0$, 
\begin{eqnarray}
I_{\rm TDE} \sim \pi-2\left(\frac{r_{\rm t}}{a_0(1-e_0^2)}\right)^{1/4}\,.
\end{eqnarray}
The bottom panel of Figure~\ref{fig:disks} shows $\pi-I_{\rm TDE}$ as a function of $a_0(1-e_0^2)\sim r$. Stars with orbital inclination $i>I_{\rm TDE}$ and $i<\pi-h$ can be tidally disrupted by the SMBH due to the disk-star interactions. The associated timescale for this disk-induced TDE is 
\begin{eqnarray}\label{eq:t-cap}
    \tau_{\rm cap}\sim \frac{\Sigma_*}{\Sigma}T \,,
\end{eqnarray}

where $\Sigma$ is the surface density of the accretion disk, $\Sigma_*=\frac{m_*}{\pi R_*^2}$ is the surface density of the star, and $T$ is the orbital period around the SMBH. The total mass of stars within the capture region is
\small
\begin{eqnarray}
M_{\rm cap}\sim \int_{r_{\rm t}}^{r_{\rm cap}} dr \int^{\arccos(I_{\rm TDE}(r))}_{\cos(\pi-h)}2\pi r^2 \rho_{\rm NSC}(r)d\cos I\,, \label{eq:cap-budget}
\end{eqnarray}
\normalsize
where $r_{\rm cap}\sim 16r_t/h^4(r_{\rm cap})$ is the truncation radius beyond which $I_{\rm TDE}$ is below the disk aspect ratio $h$ and $\rho_{\rm NSC}$ is the stellar density in the NSC.

The disk-induced TDE rate decays in a time-dependent manner, as described by \citet{Wang2023b}. However, the ratio of CL-AGN to all AGNs depends on the average disk-induced TDE rate over the lifetime of the AGN phase. The average disk-induced TDE rate during the AGN phase can therefore be written as
\begin{eqnarray}\label{eq:ave-rate}
    \langle{\dot{N}}_{\rm AGN-TDE}\rangle\sim \frac{N_{(\tau_{\rm cap}<\tau_{\rm AGN}) | (I > I_{\rm TDE})}}{\tau_{\rm AGN}}\,,
\end{eqnarray}
where $N_{(\tau_{\rm cap}<\tau_{\rm AGN})|(I > I_{\rm TDE})}$ is the total number of stars that can be captured within the tidal radius during the lifetime of the AGN disk, denoted by $\tau_{\rm AGN}$. Given that the disk-induced TDE rate decays in a time-dependent manner, this implies that short-lived AGN disks lead to a higher averaged disk-induced TDE rate in the AGN phase.
Figure~\ref{fig:enhanced} shows the enhanced TDE rate ratio between AGN-TDEs and two-body relaxation TDEs in dormant galaxies. It is evident that massive AGNs with short-lived disks have an average TDE rate an order of magnitude higher than those in dormant galaxies.

\subsection{AGN disk models}
We used three different AGN disk models to explore the predicted rate of disk-induced AGN-TDEs: (1) the classic \citet{SS73} $\alpha$-disk model; (2) the self-gravitating $\alpha$-disk model of \citet{MacLeod2020}; and (3) the \citet{Sirko2003} (SG) model. 

The accretion rate of the first two disk models can be approximated as:
\begin{eqnarray}
\dot{M}&=&\eta \dot{M}_{\rm Edd} = 2.2\eta\left(\frac{M_{\rm \bullet}}{10^8M_\odot}\right) M_\odot\, {\rm yr}^{-1},
\end{eqnarray}
where $\dot{M}$ is the accretion rate, $\eta$ is the accretion rate efficiency, $\dot{M}_{\rm Edd}=L_{\rm Edd}/\epsilon c^2$ is the Eddington accretion rate, $L_{\rm Edd}$ is the Eddington luminosity, and $\epsilon \sim 0.1$ is a constant representing the efficiency of conversion of rest mass energy to luminosity. For the Shakura-Sunyeav $\alpha$-disk model, the accretion disk surface density ($\Sigma$) and aspect ratio ($h=H/r$, where $H$ is the disk scale height at radius $r$ from the SMBH) profiles are given by
\begin{eqnarray}
\Sigma &\sim& 3\times 10^7\, \alpha^{-4/5}\left(\frac{\dot{M}}{1M_\odot {\rm yr}^{-1}}\right)^{7/10} \left(\frac{M_\bullet}{10^8 M_\odot}\right)^{1/4}\nonumber\\
&&\left(\frac{r}{1\, {\rm AU}}\right)^{-3/4} {\rm g}\, {\rm cm}^{-2},\\
h&\sim& 1.2\times 10^{-3}\, \alpha^{-1/10}\left(\frac{\dot{M}}{1M_\odot {\rm yr}^{-1}}\right)^{3/20}\left(\frac{M_\bullet}{10^8 M_\odot}\right)^{-3/8}\nonumber\\
&&\left(\frac{r}{1\, {\rm AU}}\right)^{1/8}\,,
\end{eqnarray}
where $\alpha$ is the viscosity parameter.
For the self-gravitating $\alpha$ disk, the disk surface density and aspect ratio are
\begin{eqnarray}
\Sigma &=& \frac{\dot{M}}{2\pi r v_r},\\
h &=&  \left(\frac{Q\dot{M}}{2\alpha M_{\rm \bullet}\Omega}\right)^{1/3},
\end{eqnarray}
where
\begin{eqnarray}
v_r &=& \alpha h^2 \left(\frac{GM_{\rm \bullet}}{r}\right)^{1/2},
\end{eqnarray}
$v_r$ is the radial velocity of the disk material, $Q$ is the Toomre $Q$ parameter, and $\Omega = \sqrt{GM_{\rm \bullet}/r^3}$ is the orbital frequency. We adopt $Q=\eta = 1$ in this work.

The SG model is designed to fit the `big blue bump' in AGN spectral energy distributions (SEDs) caused by inner disk luminosity. This model is gravitationally stable in the inner disk but unstable in outer regions. 

\begin{figure}
    \includegraphics[width=\columnwidth]{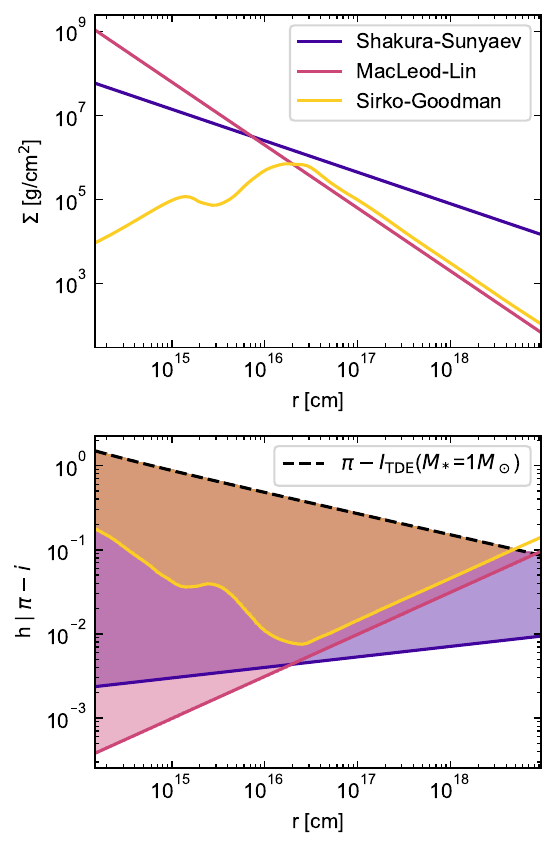}
    \caption{The three different AGN accretion disk models considered here, evaluated around a $10^8$ $M_\odot$ SMBH (see \S~2.2 for details). The upper panel shows the disk surface density ($\Sigma$) while the bottom panel shows the disk aspect ratio ($h$) and stellar orbit inclination ($i$). The shaded regions in the bottom panel indicate the disk-capture TDE region, while in-disk TDEs occur in the regions below the disk aspect ratio.}
    \label{fig:disks}
\end{figure}

Fig.~\ref{fig:disks} compares the surface density and aspect ratio of the three disk models around a $10^8 M_\odot$ SMBH. We adopt an $\alpha$ parameter for the Shakura-Sunyeav and self-gravitating disk of 0.1. In the inner region, the surface density of the $\alpha$ disk models is significantly higher than in the SG model.

\subsection{Nuclear star cluster models}

Disk-induced TDEs primarily occur in the inner regions of the disk, where $r < r_{\rm cap}$ and $I > I_{\rm TDE}$. The rate of disk-induced TDEs strongly depends on both the disk's surface density and the stellar populations in these inner regions. 

In inactive galaxies, TDEs due to $\mathcal{O}(1)M_\odot$ mass stars occur around non-spinning SMBHs ($a = 0$) only if $M_{\rm \bullet} \leq 10^8M_\odot$. However, TDEs can still occur for $M_{\rm \bullet}> 10^{8}M_{\odot}$ for more massive stars and/or if $a > 0$. Therefore, we adopt a multi-mass NSC model, where the NSC follows an IMF characterized by a power-law index $\gamma_{\rm IMF}$, where the number of stars of mass $m_*$ scales as $m_*^{-\gamma_{\rm IMF}}$. For a multi-mass, cuspy NSC, \cite{Bahcall1977} determined the equilibrium distribution:
\begin{eqnarray}
&&\rho_i(r) \propto r^{\beta_i}\label{eq:profile}\\
&&\frac{\beta_i + 3/2}{m_i} = {\rm const}\,,
\end{eqnarray}
where $\beta_i$ is the radial density power-law index corresponding to stars of mass $m_i$. For the most massive objects in the NSC, $\beta_i = -7/4$, whereas for the lightest objects, $\beta_i \sim -3/2$. The segregation timescale for stars with mass $m_i$ to reach equipartition is given by \citet{Spitzer1969}:
\begin{eqnarray}
\tau_{\rm ms,i} &\sim& \frac{m_{\rm min}}{m_i}\tau_{\rm rlx}\\
&\sim& \frac{m_{\rm min}}{m_i}\frac{\sqrt{2}\pi}{64}\frac{\sigma^3}{\ln\Lambda n G^2 \langle m \rangle^2}\\
&\sim& 3.67\times10^8\, {\rm yr}\, \frac{m_{\rm min}}{m_i} \left(\frac{\ln \Lambda}{10}\right)^{-1}\left(\frac{\sigma}{100\, {\rm km}\, {\rm s}^{-1}}\right)^{3}\nonumber\\
&&\left(\frac{n}{10^6\, {\rm pc}^{-3}}\right)^{-1}\left(\frac{\langle m \rangle}{1M_\odot}\right)^{-2},
\end{eqnarray}
where $m_{\rm min}$ is the minimum stellar mass in the NSC, $\tau_{\rm rlx}$ is the two-body relaxation timescale, $\sigma$ is the velocity dispersion of the NSC, $\ln \Lambda$ is the Coulomb logarithm, $n$ is the number density of stars in the NSC, and $\langle m_i \rangle$ is the average stellar mass within the NSC. {This timescale implies that massive stars in the dense core region of the NSC can be sufficiently segregated while the lighter stars in the outskirts of the NSC are weakly segregated.}

We assume the total NSC stellar mass within the gravitational influence radius $r_h$ around the SMBH is approximately $2M_{\rm \bullet}$ \citep{Merritt2013}, where
$r_h \sim \frac{GM_{\rm \bullet}}{\sigma^2}$. The mass of the zero-age-main-sequence (ZAMS) star ranges from $0.2\, M_\odot$ to $120\, M_\odot$. During the disk-induced TDE process, ZAMS stars follow simple stellar evolution with their stellar lifetime given by \citet{harwit2012}:
\begin{equation}
\tau_{\rm ZAMS}\sim 10\left(\frac{M_\odot}{m_i}\right)^{5/2} \rm Gyr\,.
\end{equation}
Massive stars with $\tau_{\rm ZAMS} < \tau_{\rm cap}$ evolve into compact objects before being captured within the SMBH tidal radius and will not contribute to $N_{(\tau_{\rm cap} < \tau_{\rm AGN})|(I > I_{\rm TDE})}$ in the averaged TDE rate calculations. For main-sequence stars, we adopt the mass-radius relationship of \citet{Kippenhahn2013} for $r_t$ calculations:
\begin{equation}
R_i = \left(\frac{m_i}{M_\odot}\right)^{4/5}R_\odot
\end{equation}
where $R_i$ represents the stellar radius of stars of mass $m_i$.

\subsection{CL-AGN selection criteria}

There are approximately 700 optical CL-AGN reported in the literature to date, ranging from single object discoveries to the results of systematic searches. A variety of selection criteria have been employed, with some selections based solely on photometric variability above some threshold over some timescale \citep[e.g.,][]{MacLeod2016}, while others require spectroscopic changes \citep[e.g.,][]{Stern2017}.  However, a broadly consistent sample can be constructed from a subset which employ a similar set of spectroscopic constraints in the following papers: \citet{Yang2018, MacLeod2019, Graham2020, Guo2020, Hon2022}. These are based on measurements of the H$\beta$ emission line and so we restrict our sample to $z < 0.85$ to ensure that H$\beta$ is covered by most optical spectra. Single-epoch virial black hole mass estimates were obtained from the SDSS DR18 AGN sample of \cite{Wu2022} and we consider all 93,394 AGN at $z < 0.85$ as the parent population. Our final sample comprises 66 turn-on and 100 turn-off CL-AGNs. 

SMBH mass estimates have typical uncertainties of a few tenths dex (i.e., $\Delta \log M_\bullet \sim 0.3 - 0.4$). We account for this in constructing the binned mass distributions of the samples by treating each individual mass as a Gaussian random variable with mean equal to the estimated virial mass and standard deviation equal to the uncertainty on the mass. The contribution of a source with a mass $m$ and uncertainty $\sigma$ to a histogram bin with lower and upper limits $(a, b)$ is given by:
\begin{equation}
c =  \frac{1}{2} \left[ {\rm erf}\left(\frac{b - x}{\sqrt{2}\sigma}\right) -  {\rm erf}\left(\frac{a - x}{\sqrt{2}\sigma}\right) \right],
\end{equation}
\noindent
where ${\rm erf}$ is the error function. The variance on the total bin value is: 
\begin{equation}
{\rm var}(B) = \sum_{i = 1}^{N}c_i(B)(1-c_i(B)).
\end{equation}

\begin{figure}
    \includegraphics[width=\columnwidth]{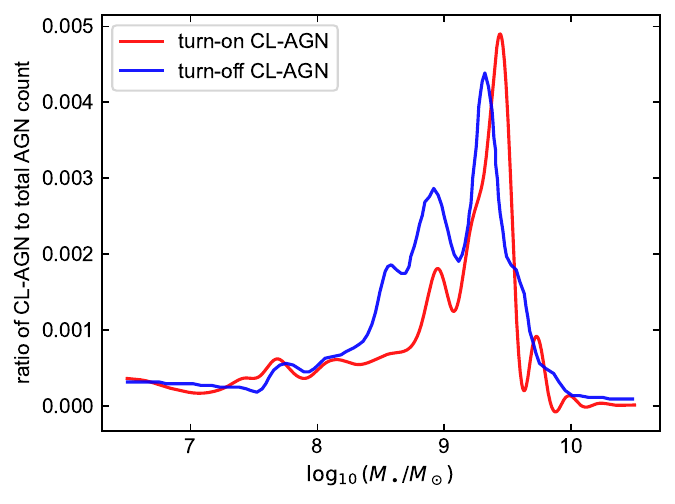}
    \caption{The single-epoch virial mass distribution of turn-on (red) and turn-off (blue) CL-AGN. The masses are drawn from \cite{Wu2022}. Uncertainties on the individual masses have been accounted for (see text for details).}
    \label{fig:cl-agn-fraction}
\end{figure}

Figure~\ref{fig:cl-agn-fraction} shows the ratio of turn-on and turn-off AGN to the total number of AGN $f_{\rm CL-AGN}$ as a function of SMBH mass. The data indicates that the ratio for both turn-on and turn-off AGN peaks at $\sim 10^9 M_\odot$ SMBHs, followed by a sharp exponential decay. This trend is consistent with the AGN-TDE rate reported in \citet{Wang2024}, if SMBHs more massive than $10^{8}M_{\odot}$ are generally rapidly spinning and as more massive stars are more efficiently captured and transported by the AGN disk due to disk-star interactions.

\section{Results}\label{sec:results}
 
\subsection{AGN-TDE rate as a function of SMBH mass}
Balmer emission line and {\it Swift}/XRT X-ray monitoring of a few repeating CL-AGN suggest a turn-on timescale of approximately one year \citep{Parker2016, Mathur2018, WangJ2023}, which aligns with the typical TDE timescale \citep{Rees1988}.  This similarity in timescales implies that if TDEs trigger turn-on CL-AGNs, the proportion of turn-on CL-AGNs among all AGNs should correspond to the TDE rate within AGNs.

\begin{figure*}
    \includegraphics[width=2\columnwidth]{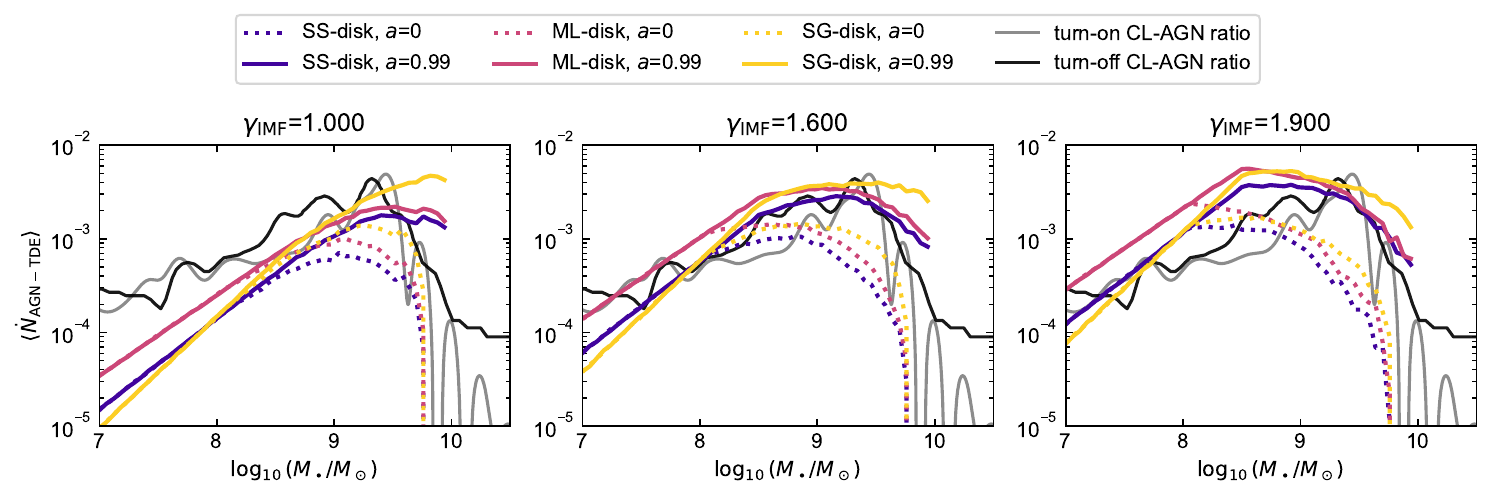}
    \caption{Disc captured+relaxation TDE rate as a function of SMBH mass. The three panels show different stellar IMF power-law indices. In each panel, the grey and black { solid} curves show the observed fraction of turn-on and turn-off CL-AGN, respectively, relative to all AGNs as a function of $M_{\bullet}$ (Fig.~3).  The dotted lines are for TDEs around a non-spinning SMBH ($a = 0$), {  for which only stars more massive than $1\, M_{\odot}$ are disrupted} and the solid lines are for a maximally spinning SMBH ($a = 0.99$). The assumed AGN accretion disk lifetime is $\tau_{\rm AGN}=3 \times 10^7$ yr. SS corresponds to the \citet{SS73} disk model, ML corresponds to the \citet{MacLeod2020} disk model, and SG corresponds to the \citet{Sirko2003} disk model.}
    \label{fig:tde-rate}
\end{figure*}

It is demonstrated that the disk-induced AGN-TDE rate declines as the disk-induced loss cone depletes on a timescale of $t_{\rm dep}$\citep{Wang2024}. If $\tau_{\rm AGN}>t_{\rm dep}$, the overall AGN-TDE rate reverts to the background TDE rate. Conversely, if $\tau_{\rm AGN}<t_{\rm dep}$, disk-induced AGN-TDEs will cease over time. The smaller  $\tau_{\rm AGN}$ becomes, the smaller the amount of time over which the AGN-TDE rate is elevated. If disk-induced AGN-TDEs represent a significant fraction of  CL-AGN, the frequency of CL-AGN constrains: (1) the NSC profile within the disk-capture AGN-TDE loss cone; (2) the AGN accretion disk density profile within the disk-capture TDE loss cone; and (3) the AGN lifetime, $\tau_{\rm AGN}$.

We calculated the average disk-induced AGN-TDE rate (Equation~\ref{eq:ave-rate}) alongside the steady background relaxation TDE rate around non-spinning ($a=0$) and maximally spinning ($a=0.99$) SMBHs with a range of $M_{\rm \bullet}$, assuming an AGN lifetime $\tau_{\rm AGN}=[1, 1000]$~Myr and using the NSC and accretion disk models from \S~\ref{sec:method}. Figure~\ref{fig:tde-rate} shows the AGN-TDE rates for different disk models for a range of $\gamma_{\rm IMF}$ values as a function of $M_{\bullet}$, compared to the observed rate of turn-on and turn-off CL-AGN relative to the total number of AGN.  

From Fig.~\ref{fig:tde-rate}, the total AGN-TDE rate increases with $M_{\bullet}$ until an inflexion point between $10^{8}-10^{9}M_\odot$, at which point lower mass stars start to become swallowed whole by the SMBH before being tidally disrupted. For rapidly spinning SMBHs, the inflexion point shifts to higher mass, as expected. At low SMBH masses, the $\alpha$ disks yield the highest average TDE rate, while SG disks, which have lower surface density in the inner region, produce the lowest total TDE rate. From Fig.~\ref{fig:tde-rate}, the $\alpha$-disk model for maximally spinning ($a \sim 0.99$) SMBHs produces the best agreement with CL-AGN observations.

\subsection{Constraints on AGN lifetime and NSC profile}

We simulated the AGN-TDE rate across a multi-dimensional grid spanning AGN disk lifetimes ($\tau_{\rm AGN}=[1,1000]$~Myr), NSC IMF index ($\gamma_{\rm IMF}=[0.5,2.0]$), SMBH mass ($\log (M_{\bullet}/M_\odot)=[6.5,10.2]$), three AGN disk models, and assuming both non-spinning ($a=0$) and maximally spinning ($a=0.99$) SMBHs. We then carried out $\chi^{2}$ tests comparing the resulting AGN-TDE rate as a function of SMBH mass with observations.

\begin{figure*}
\includegraphics[width=0.66\columnwidth]{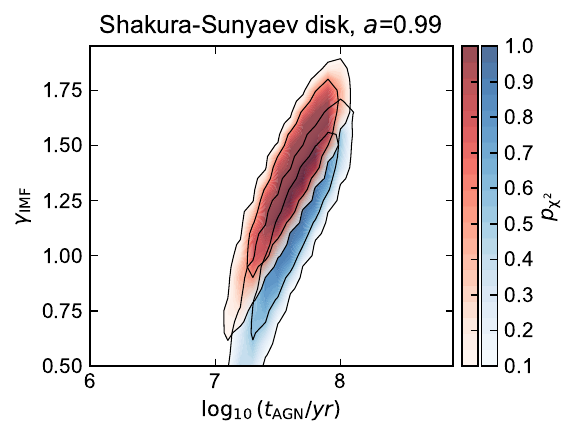}
\includegraphics[width=0.66\columnwidth]{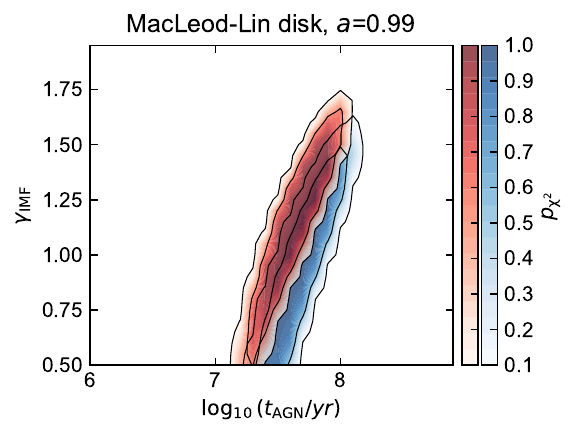}
\centering
\includegraphics[width=0.66\columnwidth]{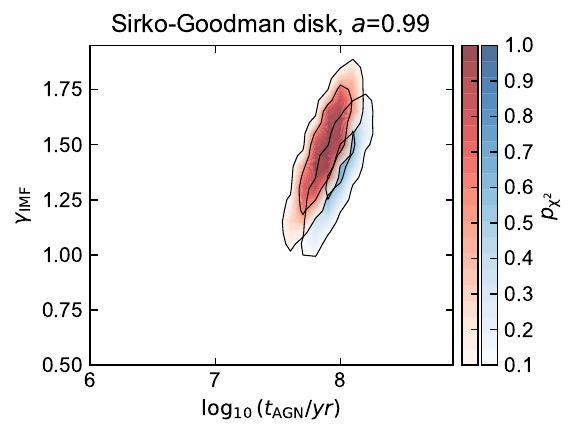}
    \caption{p-value contours from $\chi^2$ tests comparing observed CL-AGN rates to the disk-capture AGN-TDE scenario as a function of AGN disk lifetime $\tau_{\rm AGN}$ and IMF power-law index $\gamma_{\rm IMF}$. Three accretion disk models are considered (see Fig.~4). The red contours indicate turn-off CL-AGNs and the blue contours indicate turn-on CL-AGNs.}
    \label{fig:test}
\end{figure*}

Figure~\ref{fig:test} shows p-values from our $\chi^2$ tests for the three disk models. {  If we assume all CL-AGN are produced by TDEs,} the data prefer cuspy NSCs characterized by an IMF power-law index of $\gamma_{\rm IMF} \sim 1.6$ and modestly long AGN disk lifetimes ($\tau_{\rm AGN} \sim 30$~Myr). Evidently, a strong association between CL-AGN and AGN-TDEs \emph{requires} high surface-density disk models. Disk models with lower inner disk surface density can only support a much lower fraction of CL-AGN as AGN-TDEs { ($\langle{\dot{N}}_{\rm AGN-TDE}\rangle/f_{\rm CL-AGN} \sim 0.05$)}, with correspondingly weaker constraints on $\gamma_{\rm IMF}$ and $\tau_{\rm AGN}$. One possibility is that the inner disk surface density is built-up by multiple, repeated AGN-TDEs, {particularly early on in the AGN lifetime}, where most debris is captured and bound \citep{Ryu2023}. If there is a strong association between CL-AGNs and AGN-TDEs, non-spinning SMBH cases are completely ruled out by our $\chi^2$ tests. Notably, both observations and cosmological simulations find a predominance of high-spin SMBHs over a range of SMBH mass \citep[e.g.,][]{Piotrowska2024}.

\subsection{Caveats}

Our results constraining  AGN parameters rests on a plausible association between AGN-TDEs and CL-AGN, but are subject to several important possible selection effects. In particular, for all three disk models, we find best-fit AGN lifetimes on the long side of some current estimates \citep[e.g.,][]{Schawinski15,Oppenheimer18}. For the parameter constraints in Figure~\ref{fig:test}, we have assumed that all CL-AGN are due to TDEs (i.e., $\langle{\dot{N}}_{\rm AGN-TDE}\rangle/f_{\rm CL-AGN} = 1$). However, it is possible for the changing-look phenomena to have other triggers, in which case AGN disks should be much less dense than assumed in the SS, ML, or SG models, and could be more consistent with the TQM model---such a circumstance ({ $\langle{\dot{N}}_{\rm AGN-TDE}\rangle/f_{\rm CL-AGN} < 1$}) would also imply a much shorter AGN lifetime than the models in Figure~\ref{fig:test}.

On the other hand, it is also possible that there are TDEs occurring in AGN which do not trigger a substantial change in the AGN flux and do not therefore produce a CL-AGN. A fully embedded (co-planar) retrograde orbiter should produce infall of the innermost $\sim$few hundred $r_{\rm g}$ of an AGN disk \citep{Ryu2024}. By contrast, a TDE on an intersecting orbit may only produce infall of a very small region of the inner disk, $\sim 20~r_{\rm g}$ \citep[e.g.][]{Chan2019, Chan2020, Chan2021}. Notably, a few quasars have been identified where the UV emission has collapsed for a period of time \citep{Guo2016, Ross2018}. This requires the suppression or destruction of disk flux out to a few hundred $r_{\rm g}$. We could set better limits on the fraction of AGN-TDEs causing CL-AGN with a parameter space study, including detailed modeling of expected light curves for AGN-TDEs at various inclinations, semi-major axes, $M_{\bullet}$, etc. These models could serve as templates that could be compared to observed AGN light curves, setting more stringent limits on the parameters we measure.
\section{Discussion and Conclusions}\label{sec:con}

{  Assuming all CL-AGN are caused by AGN-TDEs yields good agreement with the rate of CL-AGN as a function of SMBH mass (see Fig.~\ref{fig:tde-rate}) if all SMBHs are rapidly spinning and surrounded by high surface density AGN disks embedded in NSCs with considerable mass segregation.}

A high surface density accretion disk, particularly in the innermost regions, may itself be a marker of a runaway process of stellar disk capture and AGN-TDEs. The disruption of $\mathcal{O}(10^{2})$ Sun-like stars around a $10^{8}M_{\odot}$ SMBH could add a surface density of $\sim 10^{5}\, {\rm g}\, {\rm cm}^{-2}$ to the inner $100\, r_{g}$ assuming rapid mixing of debris, as found by \citet{Ryu2023}.

X-ray reflection studies of AGN in the nearby universe reveal that most AGN with spin measurements have very high spin values \citep{Reynolds21}.  This is also supported by cosmological simulations of galaxy evolution that include SMBH feedback \citep{Piotrowska2024}. Equating CL-AGN with AGN-TDEs is consistent with this result, since Fig.~\ref{fig:tde-rate} implies that for TDEs in AGN to be responsibile for most CL-AGN, the host SMBHs must be near maximal spin ($a>0.9$). This supports a model of SMBH growth primarily by gas accretion in a consistent plane, not from chaotic accretion \citep[e.g., scenario (ii) in][]{Berti08}. This in turn suggests that AGN are fueled by occasional pulses of low-angular momentum gas accretion from long-lived, massive reservoirs.

Recent work finds that stars embedded in AGN accretion disks should rapidly accrete to  masses of $O(10^{2})M_{\odot}$ and that they may not collapse while they remain embedded in the disk \citep[e.g.,][]{Cantiello21, Jermyn22}. A range of rich dynamical encounters between these stars and other objects, including intermediate mass black holes \citep{Leigh18,McK20}, could potentially fill the AGN loss cone and generate high mass TDEs which might account for some turn-on jetted AGN \citep{Readhead24}.

We show that the observed SMBH mass distribution for turn-on CL-AGNs peaks at $\log (M_\bullet/M_\odot) \sim 8.7$. If the fraction of turn-on CL-AGNs caused by AGN-TDEs is high ($\langle{\dot{N}}_{\rm AGN-TDE}\rangle/f_{\rm CL-AGN} \sim 1$), this implies: (1) most SMBHs in CL-AGN are near maximal spin ($a>0.9$); (2) the inner AGN accretion disk surface density is high ($\geq 10^{7}\,{\rm g}\, {\rm cm}^{-2}$); (3) most AGN lifetimes are in the range $\sim 10-100$~Myr; (4) an NSC mass function that scales as $\sim m_*^{-1.6}$ is preferred, i.e. implying some mass segregation. 
{  Our results are less constraining on $\tau_{\rm AGN}$, $a$, and $\gamma_{\rm IMF}$ if only a small fraction of CL-AGN are caused by AGN-TDEs.}
Future observations of turn-on CL-AGN will help constrain the fraction CL-AGN due to AGN-TDEs, SMBH spins, AGN lifetimes, and the NSC IMF. 



\begin{acknowledgments}

YW is supported by Nevada Center for Astrophysics. KESF and BM are supported by NSF AST-2206096 and NSF AST-1831415 and Simons Foundation Grant 533845 as well as Simons Foundation sabbatical support. The Flatiron Institute is supported by the Simons Foundation. The work of DS was carried out at the Jet Propulsion
Laboratory, California Institute of Technology, under a contract with NASA. 

\end{acknowledgments}

\bibliography{sample631}{}

\begin{thebibliography}{}
\expandafter\ifx\csname natexlab\endcsname\relax\def\natexlab#1{#1}\fi
\providecommand{\url}[1]{\href{#1}{#1}}
\providecommand{\dodoi}[1]{doi:~\href{http://doi.org/#1}{\nolinkurl{#1}}}
\providecommand{\doeprint}[1]{\href{http://ascl.net/#1}{\nolinkurl{http://ascl.net/#1}}}
\providecommand{\doarXiv}[1]{\href{https://arxiv.org/abs/#1}{\nolinkurl{https://arxiv.org/abs/#1}}}

\bibitem[{{Artymowicz} {et~al.}(1993){Artymowicz}, {Lin}, \&
  {Wampler}}]{artymowicz1993}
{Artymowicz}, P., {Lin}, D.~N.~C., \& {Wampler}, E.~J. 1993, \apj, 409, 592,
  \dodoi{10.1086/172690}

\bibitem[{{Bahcall} \& {Wolf}(1977)}]{Bahcall1977}
{Bahcall}, J.~N., \& {Wolf}, R.~A. 1977, \apj, 216, 883, \dodoi{10.1086/155534}

\bibitem[{{Berti} \& {Volonteri}(2008)}]{Berti08}
{Berti}, E., \& {Volonteri}, M. 2008, \apj, 684, 822, \dodoi{10.1086/590379}

\bibitem[{{Blanchard} {et~al.}(2017){Blanchard}, {Nicholl}, {Berger},
  {Guillochon}, {Margutti}, {Chornock}, {Alexander}, {Leja}, \&
  {Drout}}]{Blanchard2017}
{Blanchard}, P.~K., {Nicholl}, M., {Berger}, E., {et~al.} 2017, \apj, 843, 106,
  \dodoi{10.3847/1538-4357/aa77f7}

\bibitem[{{Cantiello} {et~al.}(2021){Cantiello}, {Jermyn}, \&
  {Lin}}]{Cantiello21}
{Cantiello}, M., {Jermyn}, A.~S., \& {Lin}, D. N.~C. 2021, \apj, 910, 94,
  \dodoi{10.3847/1538-4357/abdf4f}

\bibitem[{{Chan} {et~al.}(2020){Chan}, {Piran}, \& {Krolik}}]{Chan2020}
{Chan}, C.-H., {Piran}, T., \& {Krolik}, J.~H. 2020, \apj, 903, 17,
  \dodoi{10.3847/1538-4357/abb776}

\bibitem[{{Chan} {et~al.}(2021){Chan}, {Piran}, \& {Krolik}}]{Chan2021}
---. 2021, \apj, 914, 107, \dodoi{10.3847/1538-4357/abf0a7}

\bibitem[{{Chan} {et~al.}(2019){Chan}, {Piran}, {Krolik}, \&
  {Saban}}]{Chan2019}
{Chan}, C.-H., {Piran}, T., {Krolik}, J.~H., \& {Saban}, D. 2019, \apj, 881,
  113, \dodoi{10.3847/1538-4357/ab2b40}

\bibitem[{{Elitzur}(2012)}]{Elitzur2012}
{Elitzur}, M. 2012, \apjl, 747, L33, \dodoi{10.1088/2041-8205/747/2/L33}

\bibitem[{{Elitzur} {et~al.}(2014){Elitzur}, {Ho}, \& {Trump}}]{Elitzur2014}
{Elitzur}, M., {Ho}, L.~C., \& {Trump}, J.~R. 2014, \mnras, 438, 3340,
  \dodoi{10.1093/mnras/stt2445}

\bibitem[{{Eracleous} {et~al.}(1995){Eracleous}, {Livio}, {Halpern}, \&
  {Storchi-Bergmann}}]{Eracleous1995}
{Eracleous}, M., {Livio}, M., {Halpern}, J.~P., \& {Storchi-Bergmann}, T. 1995,
  \apj, 438, 610, \dodoi{10.1086/175104}

\bibitem[{{Fabj} {et~al.}(2020){Fabj}, {Nasim}, {Caban}, {Ford}, {McKernan}, \&
  {Bellovary}}]{Fabj2020}
{Fabj}, G., {Nasim}, S.~S., {Caban}, F., {et~al.} 2020, \mnras, 499, 2608,
  \dodoi{10.1093/mnras/staa3004}

\bibitem[{{Generozov} \& {Perets}(2023)}]{Generozov2023}
{Generozov}, A., \& {Perets}, H.~B. 2023, \mnras, 522, 1763,
  \dodoi{10.1093/mnras/stad1016}

\bibitem[{{Graham} {et~al.}(2020){Graham}, {Ross}, {Stern}, {Drake},
  {McKernan}, {Ford}, {Djorgovski}, {Mahabal}, {Glikman}, {Larson}, \&
  {Christensen}}]{Graham2020}
{Graham}, M.~J., {Ross}, N.~P., {Stern}, D., {et~al.} 2020, \mnras, 491, 4925,
  \dodoi{10.1093/mnras/stz3244}

\bibitem[{{Guo} {et~al.}(2016){Guo}, {Malkan}, {Gu}, {Li}, {Prochaska}, {Ma},
  {You}, {Zafar}, \& {Liao}}]{Guo2016}
{Guo}, H., {Malkan}, M.~A., {Gu}, M., {et~al.} 2016, \apj, 826, 186,
  \dodoi{10.3847/0004-637X/826/2/186}

\bibitem[{{Guo} {et~al.}(2020){Guo}, {Shen}, {He}, {Wang}, {Liu}, {Wang},
  {Sun}, {Yang}, {Kong}, \& {Sheng}}]{Guo2020}
{Guo}, H., {Shen}, Y., {He}, Z., {et~al.} 2020, \apj, 888, 58,
  \dodoi{10.3847/1538-4357/ab5db0}

\bibitem[{Harwit(2012)}]{harwit2012}
Harwit, M. 2012, Astrophysical concepts (Springer Science \& Business Media)

\bibitem[{{Heckman} \& {Best}(2014)}]{Heckman14}
{Heckman}, T.~M., \& {Best}, P.~N. 2014, \araa, 52, 589,
  \dodoi{10.1146/annurev-astro-081913-035722}

\bibitem[{{Hinkle} {et~al.}(2024){Hinkle}, {Shappee}, {Auchettl}, {Kochanek},
  {Neustadt}, {Polin}, {Strader}, {Holoien}, {Huber}, {Tucker}, {Ashall}, {de
  Jaeger}, {Desai}, {Do}, {Hoogendam}, \& {Payne}}]{Hinkle2024}
{Hinkle}, J.~T., {Shappee}, B.~J., {Auchettl}, K., {et~al.} 2024, arXiv
  e-prints, arXiv:2405.08855, \dodoi{10.48550/arXiv.2405.08855}

\bibitem[{{Hon} {et~al.}(2022){Hon}, {Wolf}, {Onken}, {Webster}, \&
  {Auchettl}}]{Hon2022}
{Hon}, W.~J., {Wolf}, C., {Onken}, C.~A., {Webster}, R., \& {Auchettl}, K.
  2022, \mnras, 511, 54, \dodoi{10.1093/mnras/stab3694}

\bibitem[{{Hutsem{\'e}kers} {et~al.}(2019){Hutsem{\'e}kers}, {Ag{\'\i}s
  Gonz{\'a}lez}, {Marin}, {Sluse}, {Ramos Almeida}, \& {Acosta
  Pulido}}]{Hutsemekers2019}
{Hutsem{\'e}kers}, D., {Ag{\'\i}s Gonz{\'a}lez}, B., {Marin}, F., {et~al.}
  2019, \aap, 625, A54, \dodoi{10.1051/0004-6361/201834633}

\bibitem[{{Jermyn} {et~al.}(2022){Jermyn}, {Dittmann}, {McKernan}, {Ford}, \&
  {Cantiello}}]{Jermyn22}
{Jermyn}, A.~S., {Dittmann}, A.~J., {McKernan}, B., {Ford}, K.~E.~S., \&
  {Cantiello}, M. 2022, \apj, 929, 133, \dodoi{10.3847/1538-4357/ac5d40}

\bibitem[{{Kaur} \& {Stone}(2024)}]{Kaur2024}
{Kaur}, K., \& {Stone}, N.~C. 2024, arXiv e-prints, arXiv:2405.18500,
  \dodoi{10.48550/arXiv.2405.18500}

\bibitem[{{Kippenhahn} {et~al.}(2013){Kippenhahn}, {Weigert}, \&
  {Weiss}}]{Kippenhahn2013}
{Kippenhahn}, R., {Weigert}, A., \& {Weiss}, A. 2013, {Stellar Structure and
  Evolution}, \dodoi{10.1007/978-3-642-30304-3}

\bibitem[{{Kroupa}(2002)}]{Kroupa2002}
{Kroupa}, P. 2002, Science, 295, 82, \dodoi{10.1126/science.1067524}

\bibitem[{{Kynoch} {et~al.}(2019){Kynoch}, {Ward}, {Lawrence}, {Bruce},
  {Landt}, \& {MacLeod}}]{Kynoch2019}
{Kynoch}, D., {Ward}, M.~J., {Lawrence}, A., {et~al.} 2019, \mnras, 485, 2573,
  \dodoi{10.1093/mnras/stz517}

\bibitem[{{LaMassa} {et~al.}(2015){LaMassa}, {Cales}, {Moran}, {Myers},
  {Richards}, {Eracleous}, {Heckman}, {Gallo}, \& {Urry}}]{LaMassa15}
{LaMassa}, S.~M., {Cales}, S., {Moran}, E.~C., {et~al.} 2015, \apj, 800, 144,
  \dodoi{10.1088/0004-637X/800/2/144}

\bibitem[{{Leigh} {et~al.}(2018){Leigh}, {Geller}, {McKernan}, {Ford}, {Mac
  Low}, {Bellovary}, {Haiman}, {Lyra}, {Samsing}, {O'Dowd}, {Kocsis}, \&
  {Endlich}}]{Leigh18}
{Leigh}, N.~W.~C., {Geller}, A.~M., {McKernan}, B., {et~al.} 2018, \mnras, 474,
  5672, \dodoi{10.1093/mnras/stx3134}

\bibitem[{{Levin}(2007)}]{Yuri07}
{Levin}, Y. 2007, \mnras, 374, 515, \dodoi{10.1111/j.1365-2966.2006.11155.x}

\bibitem[{{MacLeod} {et~al.}(2016){MacLeod}, {Ross}, {Lawrence}, {Goad},
  {Horne}, {Burgett}, {Chambers}, {Flewelling}, {Hodapp}, {Kaiser}, {Magnier},
  {Wainscoat}, \& {Waters}}]{MacLeod2016}
{MacLeod}, C.~L., {Ross}, N.~P., {Lawrence}, A., {et~al.} 2016, \mnras, 457,
  389, \dodoi{10.1093/mnras/stv2997}

\bibitem[{{MacLeod} {et~al.}(2019){MacLeod}, {Green}, {Anderson}, {Bruce},
  {Eracleous}, {Graham}, {Homan}, {Lawrence}, {LeBleu}, {Ross}, {Ruan},
  {Runnoe}, {Stern}, {Burgett}, {Chambers}, {Kaiser}, {Magnier}, \&
  {Metcalfe}}]{MacLeod2019}
{MacLeod}, C.~L., {Green}, P.~J., {Anderson}, S.~F., {et~al.} 2019, \apj, 874,
  8, \dodoi{10.3847/1538-4357/ab05e2}

\bibitem[{{MacLeod} \& {Lin}(2020)}]{MacLeod2020}
{MacLeod}, M., \& {Lin}, D. N.~C. 2020, \apj, 889, 94,
  \dodoi{10.3847/1538-4357/ab64db}

\bibitem[{{Mathur} {et~al.}(2018){Mathur}, {Denney}, {Gupta}, {Vestergaard},
  {De Rosa}, {Krongold}, {Nicastro}, {Collinson}, {Goad}, {Korista}, {Pogge},
  \& {Peterson}}]{Mathur2018}
{Mathur}, S., {Denney}, K.~D., {Gupta}, A., {et~al.} 2018, \apj, 866, 123,
  \dodoi{10.3847/1538-4357/aadd91}

\bibitem[{{McKernan} {et~al.}(2022){McKernan}, {Ford}, {Cantiello}, {Graham},
  {Jermyn}, {Leigh}, {Ryu}, \& {Stern}}]{McK22}
{McKernan}, B., {Ford}, K.~E.~S., {Cantiello}, M., {et~al.} 2022, \mnras, 514,
  4102, \dodoi{10.1093/mnras/stac1310}

\bibitem[{{McKernan} {et~al.}(2020){McKernan}, {Ford}, \&
  {O'Shaughnessy}}]{McK20}
{McKernan}, B., {Ford}, K.~E.~S., \& {O'Shaughnessy}, R. 2020, \mnras, 498,
  4088, \dodoi{10.1093/mnras/staa2681}

\bibitem[{{Merloni} {et~al.}(2015){Merloni}, {Dwelly}, {Salvato},
  {Georgakakis}, {Greiner}, {Krumpe}, {Nandra}, {Ponti}, \&
  {Rau}}]{Merloni2015}
{Merloni}, A., {Dwelly}, T., {Salvato}, M., {et~al.} 2015, \mnras, 452, 69,
  \dodoi{10.1093/mnras/stv1095}

\bibitem[{{Merritt}(2013)}]{Merritt2013}
{Merritt}, D. 2013, {Dynamics and Evolution of Galactic Nuclei}

\bibitem[{{Nenkova} {et~al.}(2008){Nenkova}, {Sirocky}, {Nikutta},
  {Ivezi{\'c}}, \& {Elitzur}}]{Nenkova2008}
{Nenkova}, M., {Sirocky}, M.~M., {Nikutta}, R., {Ivezi{\'c}}, {\v{Z}}., \&
  {Elitzur}, M. 2008, \apj, 685, 160, \dodoi{10.1086/590483}

\bibitem[{{Oppenheimer} {et~al.}(2018){Oppenheimer}, {Segers}, {Schaye},
  {Richings}, \& {Crain}}]{Oppenheimer18}
{Oppenheimer}, B.~D., {Segers}, M., {Schaye}, J., {Richings}, A.~J., \&
  {Crain}, R.~A. 2018, \mnras, 474, 4740, \dodoi{10.1093/mnras/stx2967}

\bibitem[{{Parker} {et~al.}(2016){Parker}, {Komossa}, {Kollatschny}, {Walton},
  {Schartel}, {Santos-Lle{\'o}}, {Harrison}, {Fabian}, {Zetzl}, {Grupe},
  {Rodr{\'\i}guez-Pascual}, \& {Vasudevan}}]{Parker2016}
{Parker}, M.~L., {Komossa}, S., {Kollatschny}, W., {et~al.} 2016, \mnras, 461,
  1927, \dodoi{10.1093/mnras/stw1449}

\bibitem[{{Piotrowska} {et~al.}(2024){Piotrowska}, {Garc{\'\i}a}, {Walton},
  {Beckmann}, {Stern}, {Ballantyne}, {Wilkins}, {Bianchi}, {Boorman},
  {Buchner}, {Chen}, {Coppi}, {Dauser}, {Fabian}, {Kammoun}, {Madsen},
  {Mallick}, {Matt}, {Matzeu}, {Nardini}, {Pizzetti}, {Puccetti}, {Ricci},
  {Tombesi}, {Torres-Alb{\`a}}, \& {Wong}}]{Piotrowska2024}
{Piotrowska}, J.~M., {Garc{\'\i}a}, J.~A., {Walton}, D.~J., {et~al.} 2024,
  Frontiers in Astronomy and Space Sciences, 11, 1324796,
  \dodoi{10.3389/fspas.2024.1324796}

\bibitem[{{Readhead} {et~al.}(2024){Readhead}, {Ravi}, {Blandford}, {Sullivan},
  {Somalwar}, {Begelman}, {Birkinshaw}, {Liodakis}, {Lister}, {Pearson},
  {Taylor}, {Wilkinson}, {Globus}, {Kiehlmann}, {Lawrence}, {Murphy},
  {O'Neill}, {Pavlidou}, {Sheldahl}, {Siemiginowska}, \& {Tassis}}]{Readhead24}
{Readhead}, A.~C.~S., {Ravi}, V., {Blandford}, R.~D., {et~al.} 2024, \apj, 961,
  242, \dodoi{10.3847/1538-4357/ad0c55}

\bibitem[{{Rees}(1988)}]{Rees1988}
{Rees}, M.~J. 1988, \nat, 333, 523, \dodoi{10.1038/333523a0}

\bibitem[{{Reynolds}(2021)}]{Reynolds21}
{Reynolds}, C.~S. 2021, \araa, 59, 117,
  \dodoi{10.1146/annurev-astro-112420-035022}

\bibitem[{{Risaliti} {et~al.}(2002){Risaliti}, {Elvis}, \&
  {Nicastro}}]{Risaliti02}
{Risaliti}, G., {Elvis}, M., \& {Nicastro}, F. 2002, \apj, 571, 234,
  \dodoi{10.1086/324146}

\bibitem[{{Rivers} {et~al.}(2015){Rivers}, {Balokovi{\'c}}, {Ar{\'e}valo},
  {Bauer}, {Boggs}, {Brandt}, {Brightman}, {Christensen}, {Craig}, {Gandhi},
  {Hailey}, {Harrison}, {Koss}, {Ricci}, {Stern}, {Walton}, \&
  {Zhang}}]{Rivers15}
{Rivers}, E., {Balokovi{\'c}}, M., {Ar{\'e}valo}, P., {et~al.} 2015, \apj, 815,
  55, \dodoi{10.1088/0004-637X/815/1/55}

\bibitem[{{Ross} {et~al.}(2020){Ross}, {Graham}, {Calderone}, {Ford},
  {McKernan}, \& {Stern}}]{Ross20}
{Ross}, N.~P., {Graham}, M.~J., {Calderone}, G., {et~al.} 2020, \mnras, 498,
  2339, \dodoi{10.1093/mnras/staa2415}

\bibitem[{{Ross} {et~al.}(2018){Ross}, {Ford}, {Graham}, {McKernan}, {Stern},
  {Meisner}, {Assef}, {Dey}, {Drake}, {Jun}, \& {Lang}}]{Ross2018}
{Ross}, N.~P., {Ford}, K.~E.~S., {Graham}, M., {et~al.} 2018, \mnras, 480,
  4468, \dodoi{10.1093/mnras/sty2002}

\bibitem[{{Ruan} {et~al.}(2016){Ruan}, {Anderson}, {Cales}, {Eracleous},
  {Green}, {Morganson}, {Runnoe}, {Shen}, {Wilkinson}, {Blanton}, {Dwelly},
  {Georgakakis}, {Greene}, {LaMassa}, {Merloni}, \& {Schneider}}]{Ruan2016}
{Ruan}, J.~J., {Anderson}, S.~F., {Cales}, S.~L., {et~al.} 2016, \apj, 826,
  188, \dodoi{10.3847/0004-637X/826/2/188}

\bibitem[{{Runco} {et~al.}(2016){Runco}, {Cosens}, {Bennert}, {Scott},
  {Komossa}, {Malkan}, {Lazarova}, {Auger}, {Treu}, \& {Park}}]{Runco16}
{Runco}, J.~N., {Cosens}, M., {Bennert}, V.~N., {et~al.} 2016, \apj, 821, 33,
  \dodoi{10.3847/0004-637X/821/1/33}

\bibitem[{{Ryu} {et~al.}(2020{\natexlab{a}}){Ryu}, {Krolik}, {Piran}, \&
  {Noble}}]{Ryu2020a}
{Ryu}, T., {Krolik}, J., {Piran}, T., \& {Noble}, S.~C. 2020{\natexlab{a}},
  \apj, 904, 99, \dodoi{10.3847/1538-4357/abb3cd}

\bibitem[{{Ryu} {et~al.}(2020{\natexlab{b}}){Ryu}, {Krolik}, {Piran}, \&
  {Noble}}]{Ryu2020b}
---. 2020{\natexlab{b}}, \apj, 904, 100, \dodoi{10.3847/1538-4357/abb3ce}

\bibitem[{{Ryu} {et~al.}(2023){Ryu}, {McKernan}, {Ford}, {Cantiello}, {Graham},
  {Stern}, \& {Leigh}}]{Ryu2023}
{Ryu}, T., {McKernan}, B., {Ford}, K.~E.~S., {et~al.} 2023, \mnras,
  \dodoi{10.1093/mnras/stad3487}

\bibitem[{{Ryu} {et~al.}(2024){Ryu}, {McKernan}, {Ford}, {Cantiello}, {Graham},
  {Stern}, \& {Leigh}}]{Ryu2024}
---. 2024, \mnras, 527, 8103, \dodoi{10.1093/mnras/stad3487}

\bibitem[{{Salpeter}(1955)}]{Salpeter1955}
{Salpeter}, E.~E. 1955, \apj, 121, 161, \dodoi{10.1086/145971}

\bibitem[{{Schawinski} {et~al.}(2015){Schawinski}, {Koss}, {Berney}, \&
  {Sartori}}]{Schawinski15}
{Schawinski}, K., {Koss}, M., {Berney}, S., \& {Sartori}, L.~F. 2015, \mnras,
  451, 2517, \dodoi{10.1093/mnras/stv1136}

\bibitem[{{Shakura} \& {Sunyaev}(1973)}]{SS73}
{Shakura}, N.~I., \& {Sunyaev}, R.~A. 1973, \aap, 24, 337

\bibitem[{{Shapovalova} {et~al.}(2010){Shapovalova}, {Popovi{\'c}}, {Burenkov},
  {Chavushyan}, {Ili{\'c}}, {Kova{\v{c}}evi{\'c}}, {Bochkarev}, \&
  {Le{\'o}n-Tavares}}]{Shapovalova2010}
{Shapovalova}, A.~I., {Popovi{\'c}}, L.~{\v{C}}., {Burenkov}, A.~N., {et~al.}
  2010, \aap, 509, A106, \dodoi{10.1051/0004-6361/200912311}

\bibitem[{{Sheng} {et~al.}(2017){Sheng}, {Wang}, {Jiang}, {Yang}, {Yan}, {Dou},
  \& {Peng}}]{Sheng2017}
{Sheng}, Z., {Wang}, T., {Jiang}, N., {et~al.} 2017, \apjl, 846, L7,
  \dodoi{10.3847/2041-8213/aa85de}

\bibitem[{Sirko \& Goodman(2003)}]{Sirko2003}
Sirko, E., \& Goodman, J. 2003, Mon. Not. R. Astron. Soc, 341, 501.
\newblock \url{https://academic.oup.com/mnras/article/341/2/501/1105444}

\bibitem[{{Spitzer}(1969)}]{Spitzer1969}
{Spitzer}, Lyman, J. 1969, \apjl, 158, L139, \dodoi{10.1086/180451}

\bibitem[{{Stern} {et~al.}(2017){Stern}, {Graham}, {Arav}, {Djorgovski},
  {Chamberlain}, {Barth}, {Donalek}, {Drake}, {Glikman}, {Jun}, {Mahabal}, \&
  {Steidel}}]{Stern2017}
{Stern}, D., {Graham}, M.~J., {Arav}, N., {et~al.} 2017, \apj, 839, 106,
  \dodoi{10.3847/1538-4357/aa683c}

\bibitem[{{Stern} {et~al.}(2018){Stern}, {McKernan}, {Graham}, {Ford}, {Ross},
  {Meisner}, {Assef}, {Balokovi{\'c}}, {Brightman}, {Dey}, {Drake},
  {Djorgovski}, {Eisenhardt}, \& {Jun}}]{Stern2018}
{Stern}, D., {McKernan}, B., {Graham}, M.~J., {et~al.} 2018, \apj, 864, 27,
  \dodoi{10.3847/1538-4357/aac726}

\bibitem[{{Wang} {et~al.}(2024{\natexlab{a}}){Wang}, {Xu}, {Cao}, {Gao}, {Xie},
  \& {Wei}}]{WangJM2024}
{Wang}, J., {Xu}, D.~W., {Cao}, X., {et~al.} 2024{\natexlab{a}}, arXiv
  e-prints, arXiv:2405.10663, \dodoi{10.48550/arXiv.2405.10663}

\bibitem[{{Wang} {et~al.}(2018){Wang}, {Xu}, \& {Wei}}]{Wang2018}
{Wang}, J., {Xu}, D.~W., \& {Wei}, J.~Y. 2018, \apj, 858, 49,
  \dodoi{10.3847/1538-4357/aab88b}

\bibitem[{{Wang} {et~al.}(2023{\natexlab{a}}){Wang}, {Zheng}, {Brink}, {Xu},
  {Filippenko}, {Gao}, {Xie}, \& {Wei}}]{WangJ2023}
{Wang}, J., {Zheng}, W.~K., {Brink}, T.~G., {et~al.} 2023{\natexlab{a}}, arXiv
  e-prints, arXiv:2308.16521, \dodoi{10.48550/arXiv.2308.16521}

\bibitem[{{Wang} {et~al.}(2024{\natexlab{b}}){Wang}, {Ma}, {Wu}, \&
  {Jiang}}]{WangMY2024}
{Wang}, M., {Ma}, Y., {Wu}, Q., \& {Jiang}, N. 2024{\natexlab{b}}, \apj, 960,
  69, \dodoi{10.3847/1538-4357/ad0bfb}

\bibitem[{{Wang} {et~al.}(2024{\natexlab{c}}){Wang}, {Lin}, {Zhang}, \&
  {Zhu}}]{Wang2024}
{Wang}, Y., {Lin}, D. N.~C., {Zhang}, B., \& {Zhu}, Z. 2024{\natexlab{c}},
  \apjl, 962, L7, \dodoi{10.3847/2041-8213/ad20e5}

\bibitem[{{Wang} {et~al.}(2023{\natexlab{b}}){Wang}, {Zhu}, \&
  {Lin}}]{Wang2023b}
{Wang}, Y., {Zhu}, Z., \& {Lin}, D. N.~C. 2023{\natexlab{b}}, arXiv e-prints,
  arXiv:2308.09129, \dodoi{10.48550/arXiv.2308.09129}

\bibitem[{{Wu} \& {Shen}(2022)}]{Wu2022}
{Wu}, Q., \& {Shen}, Y. 2022, \apjs, 263, 42, \dodoi{10.3847/1538-4365/ac9ead}

\bibitem[{{Wu} {et~al.}(2024){Wu}, {Chen}, \& {Lin}}]{wu2024}
{Wu}, Y., {Chen}, Y.-X., \& {Lin}, D. N.~C. 2024, \mnras, 528, L127,
  \dodoi{10.1093/mnrasl/slad183}

\bibitem[{{Yang} {et~al.}(2018){Yang}, {Wu}, {Fan}, {Jiang}, {McGreer},
  {Shangguan}, {Yao}, {Wang}, {Joshi}, {Green}, {Wang}, {Feng}, {Fu}, {Yang},
  \& {Liu}}]{Yang2018}
{Yang}, Q., {Wu}, X.-B., {Fan}, X., {et~al.} 2018, \apj, 862, 109,
  \dodoi{10.3847/1538-4357/aaca3a}

\bibitem[{{Zhou} {et~al.}(2024){Zhou}, {Huang}, {Guo}, {Li}, \&
  {Pan}}]{Zhou2024}
{Zhou}, C., {Huang}, L., {Guo}, K., {Li}, Y.-P., \& {Pan}, Z. 2024, arXiv
  e-prints, arXiv:2401.11190, \dodoi{10.48550/arXiv.2401.11190}

\bibitem[{{Zhou} {et~al.}(2022){Zhou}, {Deng}, {Chen}, \& {Lin}}]{Zhou2022}
{Zhou}, T., {Deng}, H.-P., {Chen}, Y.-X., \& {Lin}, D. N.~C. 2022, \apj, 940,
  117, \dodoi{10.3847/1538-4357/ac9bf6}

\end{thebibliography}
\bibliographystyle{aasjournal}



\end{document}